\documentclass[twocolumn,aps,prc,superscriptaddress,showpacs,nobibnotes]{revtex4}
\usepackage{amsmath,bm}%
\usepackage{graphicx}%
\usepackage{ulem}

\begin{document}

\draft
\title{ Isoscaling behavior in the Isospin dependent Quantum Molecular
Dynamics model }

\thanks{ Supported by the Major State Basic Research Development
Program under Contract No G2000774004, the
National Natural Science Foundation of China (NNSFC) under Grant
No 10328259 and 10135030, and the Chinese Academy of Sciences
Grant for the National Distinguished Young Scholars of NNSFC.}

\author{TIAN Wen-Dong} \thanks{Email: wdtian@sinap.ac.cn, Tel:021-59556905}
\affiliation{Shanghai Institute of Applied Physics, Chinese
Academy of Sciences, P. O. Box 800-204, Shanghai 201800}
\author{MA Yu-Gang} \thanks{Corresponding author: ygma@sinr.ac.cn}
\affiliation{Shanghai Institute of Applied Physics, Chinese
Academy of Sciences, P. O. Box 800-204, Shanghai 201800}
\author{CAI Xiang-Zhou}
\affiliation{Shanghai Institute of Applied Physics, Chinese
Academy of Sciences, P. O. Box 800-204, Shanghai 201800}
\author{CHEN Jin-Gen}
\affiliation{Shanghai Institute of Applied Physics, Chinese
Academy of Sciences, P. O. Box 800-204, Shanghai 201800}
\affiliation{Graduate School of the Chinese Academy of Sciences}
\author{CHEN Jin-Hui}
\author{FANG De-Qing}
\author{GUO Wei}
\affiliation{Shanghai Institute of Applied Physics, Chinese
Academy of Sciences, P. O. Box 800-204, Shanghai 201800}
\author{MA Chun-Wang}
\author{MA Guo-Liang} \affiliation{Shanghai Institute of
Applied Physics, Chinese Academy of Sciences, P. O. Box 800-204,
Shanghai 201800} \affiliation{Graduate School of the Chinese
Academy of Sciences}
\author{SHEN Wen-Qing}
\affiliation{Shanghai Institute of Applied Physics, Chinese
Academy of Sciences, P. O. Box 800-204,  Shanghai 201800}
\author{WANG Kun}
\author{WEI Yi-Bin}
\affiliation{Shanghai Institute of Applied Physics, Chinese
Academy of Sciences, P. O. Box 800-204, Shanghai 201800}
\affiliation{Graduate School of the Chinese Academy of Sciences}
\author{YAN Ting-Zhi}
\author{ZHONG Chen}
\author{ZUO Jia-Xu}
\affiliation{Shanghai Institute of Applied Physics, Chinese
Academy of Sciences, P. O. Box 800-204,    Shanghai 201800}
\date{\today}

\begin{abstract}
The isoscaling behavior is investigated in the frame of  Isospin
dependent Quantum Molecular Dynamics (IQMD) models. The isotopic
yields ratio $Y_2$/$Y_1$ for reactions $^{48}Ca$+$^{48}Ca$ and
$^{40}Ca+^{40}Ca$ at different entrance channels are simulated and
presented, the relationship between the isoscaling parameter and
the entrance channel is analyzed, the results show that $\alpha$
and $\beta$ reduce with the rise of incident energies and increase
with the impact parameter b, which can be attributed to the
temperature varying of the pre-fragments in different entrance
channels. The relation of $\alpha$ and symmetry-term coefficient
$C_{sym}$ reveals that the chemical potential difference
$\triangle\mu$ is sensitive to the symmetry-term coefficient
$C_{sym}$.

\end{abstract}


\maketitle

Experimental investigations of the properties of isospin
asymmetric nuclear matter are receiving increased attention due to
the possibility of creating nuclear material with appreciable
isospin asymmetry using neutron-rich beams. This is expected to
enable one to extrapolate the present understanding regarding the
effective nucleon-nucleon interaction to the unknown domain of
large isospin asymmetry. The isospin asymmetry of reaction parts
plays an important role in the evolution of nucleon exchange
processes which form the basis of many nuclear transport models
\cite{Li,Gupta,Muller,Lee,Ma1999,Bom,Latt}. It is also an
effective indicator of the degree of chemical equilibrium reached
in heavy ion reactions.

To minimize undesirable complications stemming from the sequential
decays of primary unstable fragments, it has been proposed that
isospin effects can be best studied by comparing the same
observable in two similar reactions that differ mainly in isospin
asymmetry \cite{Ma1999,Tsang01a,Tsang01b,Tsang01c}. If two
reactions, 1 and 2, have the same temperature but different
isospin asymmetry, for example, the ratio of a specific isotope
yields with neutron and proton number N and Z obtained from system
2 and system 1 have been observed to exhibit isoscaling, i.e.,
exponential dependence of the form \cite{Tsang01b}:
\begin{equation}
 R_{21}(N,Z)=\frac{Y_2(N,Z)}{Y_1(N,Z)}=Cexp(\alpha N+\beta Z).
\end{equation}
where $\alpha$ and $\beta$ are the scaling parameters and $C$ is
an overall normalization constant, with the convention of that the
neutron and proton composition of reaction 2 is more neutron-rich
than that of reaction 1. The systematization of the experimental
data in form (1) has been conformed to variety of reactions over
wide energy range and reaction systems \cite{Tsang01b,Bot,Soul}.
This scaling law occurs naturally within various reaction
processes, such as the  evaporation, deep inelastic scattering
\cite{Tsang01b}, fission \cite{Veselsky2,Wang}, and
multifragmentation \cite{Tsang01b,Bot,Ma2004}. In the framework of
different models \cite{Tsang01b}, it has been revealed that the
isotope yields ratio are not affected very much by the sequential
decays following the fast stage of the reaction, so $R_{21}(N,Z)$
deduced from the detected isotope fragments can reflect the
isotope yields ratio of the primary fragments. The isoscaling
parameters $\alpha$ and $\beta$ can be used to study the isospin
dependent properties in nuclear collisions, and this method allows
one to study the early reaction stage of the decay fragmenting
system, to understand the dependence of measured isotope
distribution on properties of hot emitter, the nuclear asymmetry
term in nuclear matter equation of state (EOS), and allows the
direct comparison between experimental measurement with
theoretical simulation.

To explore the isospin properties and the fragmentation in
dynamical nuclear collisions, we study the isoscaling phenomenon
by the  IQMD model \cite{Zhang,Wei,Liu}, which is based on the general
QMD \cite{Aich,Aich2,LiZX,Ma1995} model to include explicitly
isospin-degrees of freedom. The QMD model is classical in essence
because the time evolution of the system is determined by
classical canonical equation of motion, however, many important
quantum features are included in this prescription. It is well
known that the dynamics in heavy ion collision at intermediate
energies is mainly governed by three components: the mean fields,
two-body collision and Pauli blocking, therefore, for
isospin-dependent reaction dynamics model it is important for
these three components to include isospin degrees of freedom. In
additional, in initialization of projectile and target nuclei, the
samples of neutrons nd protons in phase space are also  treated
separately since there exists a large difference between neutron
and proton density distributions for nuclei far from the
$\beta$-stability line \cite{Zhang,Wei}.

In the IQMD model, the nuclear mean field is given by
\begin{equation}
U(\rho,\tau_z) = U^{dd} + U^{Yuk} + U^{Coul} + U^{Sym} + U^{MDI}
\end{equation}
where $U^{dd}$ is the density-dependent (Skyrme) potential,
$U^{Yuk}$ the Yukawa (surface) potential, $U^{Coul}$ the Coulomb
energy, $U^{Sym}$ the symmetry energy term and $U^{MDI}$ the
momentum dependent interaction, but in our present calculation,
$U^{MDI}$ was not included. The $U^{dd}$ can be written as
\begin{equation}
U^{dd} = a (\frac{\rho}{\rho_0}) + b (\frac{\rho}{\rho_0})^\gamma
\end{equation}
with $\rho_0 = 0.16 fm^{-3}$(the normal nuclear matter density),
$a$ = -356 MeV, $b$ = 303MeV, and $\gamma$ = 1.17 (corresponding
to the Soft EOS).
\begin{eqnarray}
U^{Yuk} = \frac{V_y}{2} \sum_{i \neq j} \frac{1}{r_{ij}} exp(Lm^2)~~~~~~~~~~~~~~~~~~~\nonumber\\
 \times [exp(mr_{ij})erfc (\sqrt{L}m - r_{ij}/\sqrt{4L})~\nonumber\\
- exp(mr_{ij})erfc(\sqrt{L}m+r_{ij}/\sqrt{4L})]
\end{eqnarray}
\begin{equation}
U^{Coul} = \frac{e^2}{4}\sum_{i\neq j}
\frac{1}{r_{ij}}(1+\tau_{iz})(1-\tau_{jz})erfc(r_{ij}/\sqrt{4L})
\end{equation}

\begin{equation}
U^{Sym} = \frac{C_{sym}}{2\rho_0}\sum_{i\neq j}
\tau_{iz}\tau_{jz}\frac{1}{(4\pi
L)^{3/2}}exp[-\frac{(r_i-r_j)^2}{4L}]
\end{equation}
with $V_y$ = -0.0024 GeV, $m$ = 0.83, and the $L$ is the so-called
Gaussian wave-packet width (here $L$ = 2.0 $fm^2$). The relative
distance $r_{ij}=|r_i-r_j|$. The $\tau_{iz}$ is the $z-$th
component of the isospin degree of freedom for the $i-$th nucleon,
which is equal to 1 and -1 for proton and neutron, respectively.
$C_{sym}$ is symmetry energy strength. 
More details on QMD
 and IQMD models are discussed in Ref \cite{Zhang,Wei,Liu,Aich2}.

\begin{figure}[th]
\includegraphics[scale=1.0]{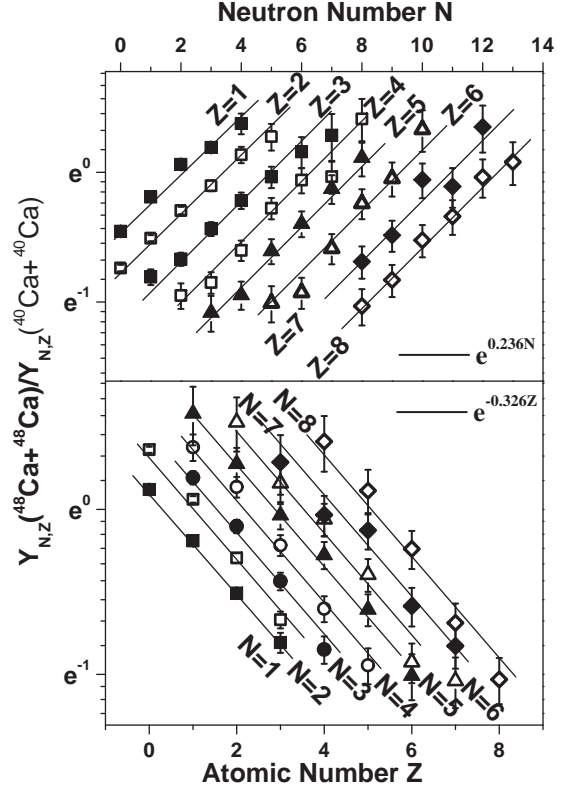}
\caption{{\protect\small {The fragment yield ratio of the IQMD
simulations for central collisions $^{48}$Ca + $^{48}$Ca and
$^{40}$Ca + $^{40}$Ca at 35MeV/A, $\alpha$ and $\beta$ are printed
inside the figure.}}}
\end{figure}

\begin{figure}[th]
\includegraphics[scale=0.7]{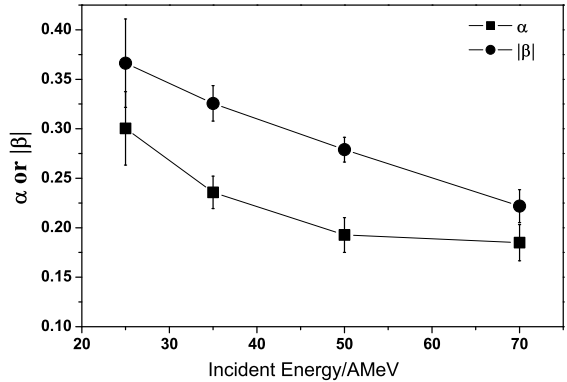}
\caption{{\protect\small {The variance of isoscaling parametes
$\alpha$ and $\beta$ with the incident energies for central
collisions $^{48}$Ca+$^{48}$Ca and $^{40}$Ca+$^{40}$Ca}}}
\end{figure}

We first perform the reaction simulations for $^{40}$Ca +
$^{40}$Ca and $^{48}$Ca + $^{48}$Ca collisions with impact
parameter b = 0 at several different incident energies $E/A$ = 25,
35, 50 and 70 MeV, the dynamical process was simulated until t =
200 fm/c. In central collisions, many fragments are formed, which
can  be basically divided into projectile-like, target-like and
neck region emitted products. In the neck region of central
collision, it is more reliable that the thermal and chemical
equilibrium could be achieved, so we select the products with the
parallel velocity condition of $|v_{//}/v_{in}| < 0.2$, where
$v_{//}$ means the parallel velocity of the products, and $v_{in}$
is the incident velocity, by which determined the fragments
emitted from the neck region parts. In the following discussion
all simulation results are under such selection.

Fig. 1 shows the yields ratio $Y_{N,Z}$($^{48}Ca +
^{48}Ca$)/$Y_{N,Z}$($^{40}Ca + ^{40}Ca$) of these two reaction at
the incident energy E/A=35MeV, clear isoscaling relations are
observed, 
the error bar inside Fig. 1 and following's plots
are statistical error only.
The isoscaling parameters $\alpha$
and $\beta$ can be extracted from formula (1), which are printed
inside the figure with $\alpha$ = 0.236, and $\beta$ = -0.326.
While extracting $\alpha$ and $\beta$, we use $R_{21}(N) =
C'exp(\alpha N)$ and $R_{21}(Z) = C^" exp(\beta Z)$, respectively.
Fig. 2 plots the variation of $\alpha$ and $\beta$ with the
incident energies for central collisions $^{40}Ca + ^{40}Ca$ and
$^{48}Ca + ^{48}Ca$ , one can find that the isoscaling parameters
$\alpha$ decreases with the increasing of the incident energy as
well as $|\beta|$. In the grand-canonical approximation
\cite{Ma1999,Tsang01b}, $\alpha = \Delta \mu_n/T$ and $\beta =
\Delta \mu_p/T$, provided a
common temperature $T$ for both systems exists \cite{Tsang01b},
$\mu_n$ and $\mu_p$ are neutron and proton chemical potential of
the system, determined by the system composite, the density
$\rho$, the temperature $T$ and its overall $N/Z$ ratio
\cite{Bot,Ma2004}. In two similar reactions like
$^{40}$Ca+$^{40}$Ca and $^{48}Ca + ^{48}Ca$, $\Delta \mu_n$ and
$\Delta \mu_p$ should  not show any significant changes with the
temperature for light charged particles as the model calculations
\cite{Bot}, therefor the reduction of $\alpha$ and $|\beta|$ with
the increasing of incident energies can be mainly attributed to
the rise of the system temperature.

\begin{figure}[th]
\includegraphics[scale=0.7]{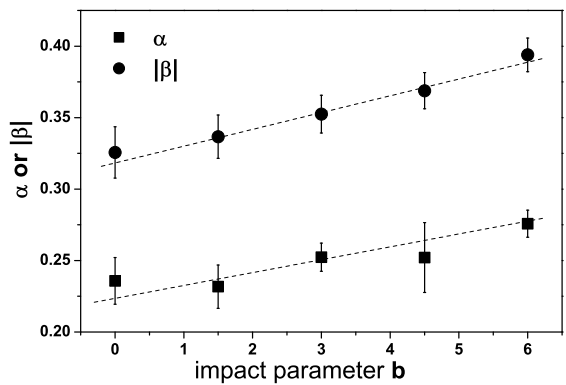}
\caption{{\protect\small {Relation of isoscaling parameters
$\alpha$ and $\beta$ with the impact parameter b, at incident
energy 35MeV/A for collisions $^{48}$Ca+$^{48}$Ca and
$^{40}$Ca+$^{40}$Ca, the dash lines are plotted to guide the eyes.
}}}
\end{figure}
\begin{figure}[th]

\includegraphics[scale=0.7]{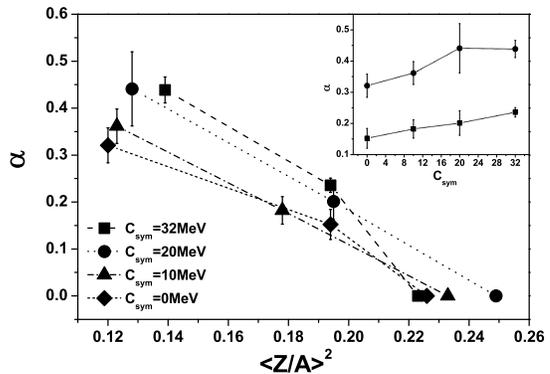}
\caption{{\protect\small {Relation between $(Z/A)^2$ and $\alpha$
for three different isospin systems $^{40}$Ca+$^{40}$Ca,
$^{48}$Ca+$^{48}$Ca and $^{60}$Ca+$^{60}$Ca with different
symmetry potential coefficient $C_{sym}$, and the insert figure is
the trend of $\alpha$ and $C_{sym}$ of two yields ratios,
the square solid point represents the result from ratio of
$Y($$^{48}$$Ca$+$^{48}$$Ca)$/$Y($$^{40}$$Ca$+$^{40}$$Ca)$, and the
round solid point is extracted from the yields ratio of
$Y($$^{60}$$Ca$+$^{60}$$Ca)$/$Y($$^{40}$$Ca$+$^{40}$$Ca)$.}}}
\end{figure}

In Fig. 3 the relation of isoscaling parameters $\alpha$ and
$|\beta|$ with the impact parameter b is shown for collision
$^{48}$Ca+$^{48}$Ca and $^{40}$Ca+$^{40}$Ca, $\alpha$ and
$|\beta|$ values raise with the increasing of b. As we know, the
reaction become less violent from central to peripheral collision,
then system  becomes less excited, the temperature $T$ of the
system generally be defined as $T$ = $\sqrt{k\langle E_0 \rangle /
\langle A_0 \rangle}$, where $E_0$ is the excitation energy of the
prefragment, and $A_0$ the mass of the decaying prefragment, $k$
denotes the inverse level density parameter, the initial
temperature $T$ determined by the excitation of the decaying
prefragment. The peripheral collision have less excitation energy
comparing with the central collision, which means low temperature
the emitting source can be reached. The free neutron and proton
chemical potential difference $\Delta \mu_n$ and $\Delta \mu_p$
does not change with the varying of the temperature very much
\cite{Bot}, the reduction of isoscaling parameters $\alpha$ and
$|\beta|$ can be caused by the temperature difference.

Furthermore, to explore the aspect of equilibrium in fragment
emission in IQMD simulation, we study the relationship between the
isoscaling and the fragment isospin asymmetry in IQMD simulation.
The following relation (7) was found relating the isoscaling
 parameter $\alpha$, the $(Z/A)^2$ of fragments, and the symmetry
  energy coefficient $C(Z)$ by the form \cite{Soul}

\begin{equation}
\alpha = 4 C_{sym}[(Z_1/A_1)^2 -(Z_2/A_2)^2] /T
\end{equation}
where $C_{sym}$ is the symmetry energy coefficient and $T$ is the
system temperature, 1 and 2 denote the neutron-deficient and
neutron-rich reaction system, respectively. In IQMD simulation,
three reactions of $^{40}$Ca+$^{40}$Ca, $^{48}$Ca+$^{48}$Ca and
$^{60}$Ca+$^{60}$Ca are simulated with different symmetry-term
potential coefficient $C_{sym}$, at the incident energy
$E/A$=35$MeV$, the relation between the $(Z/A)^2$, symmetry
potential coefficient $C_{sym}$  and the isoacaling parameter
$\alpha$ from three simulations are plotted in Fig. 4, where $Z/A$
is calculated from the emitted fragments, averaged over the
fragment $5\leq Z \leq8$. One can find that linear relation is
generally kept between $\alpha$ and $(Z/A)^2$ for different
$C_{sym}$, this linear function can prove the chemical potential
equilibration in reactions $^{40}$Ca+$^{40}$Ca,
$^{48}$Ca+$^{48}$Ca, and $^{60}$Ca+$^{60}$Ca though different
symmetry potential coefficients $C_{sym}$ are used in the
simulations, since equation (7) is satisfied only in thermal and
chemical potential equilibrium. The insert figure shows that the
isoscaling parameter $\alpha$ increases with the increasing of
symmetry-term coefficient $C_{sym}$, in both
Y($^{48}$Ca+$^{48}$Ca)/Y($^{40}$Ca+$^{40}$Ca) and
Y($^{60}$Ca+$^{60}$Ca)/Y($^{40}$Ca+$^{40}$Ca). Up to now no
evidence show asymmetry-term coefficient $C_{sym}$ have impact on
the system temperature, since the temperature of the system, as we
mentioned above, mainly determined by the incident energy, also
means the excitation energy of the reaction. Other models like
isospin dependent lattice gas model (LGM) \cite{Ma2004} and
statistical multifragmentation (SMM) model \cite{Bot}  predicted
that the chemical potential differences between two similar
reaction $\Delta \mu_n$ varies little in different temperature.
But the chemical potential difference $\Delta \mu_n$ changes a lot
with different symmetry-term coefficients $C_{sym}$, the
increasing of $\alpha$ mainly comes from the contribution of
chemical potential difference $\Delta \mu_n$.

In summary, with help of the IQMD simulation, the isotope yield
ratio in the multi-fragmentation process in intermediate energies
heavy ion collision, shows the isoscaling behavior in different
entrance channels. The isoscaling parameters $\alpha$ and $\beta$
drop with the incident energies and raise with the impact
parameters, which can be basically attributed to the different
temperature of the emitting source. The isoscaling law also
presents a  linear relation between $\alpha$ and $(Z/A)^2$, but
the $\alpha$ varies with symmetry-term coefficient $C_{sym}$ which
indicates $\alpha$ can be served as a sensitive probe to the
symmetry potential as LGM shows \cite{Ma2004}.

{}

\end{document}